\documentclass{ws-ijmpd}

\input epsf.sty

\begin{document}


%
\catchline{}{}{}{}{}
%

\title{Quark stars and quantum-magnetically induced collapse \\}

\author{A. P\'erez Mart\'{\i}nez, H. P\'{e}rez Rojas}

\address{Instituto de Cibernetica Matematica y Fisica, ICIMAF \\
Calle E esq 15 No. 309 Vedado, Havana, 10400, Cuba\\
aurora@icmf.inf.cu,hugo@icmf.inf.cu}

\author{H. J. Mosquera Cuesta}
\address{Instituto de Cosmologia, Relatividade e Astrof\'{\i}sica
(ICRA-BR),\\ Centro Brasileiro de Pesquisas F\'{\i}sicas \\ Rua
Dr. Xavier Sigaud 150, Urca, CEP 22290-180, Rio de Janeiro, Brazil\\
hermanjc@cbpf.br}

\author{M. Boligan}
\address{Centro Meteorol\'ogico de Pinar del Rio, Cuba \\
miguel.boligan@meteoro.pinar.cu}

\author{M. G. Orsaria}
\address{Nuclear Astrophysics Group, Centro Brasileiro de Pesquisas F\'{\i}sicas
\\ Rua Dr. Xavier Sigaud 150, Urca, CEP 22290-180, Rio de Janeiro, Brazil\\ orsaria@cbpf.br}

\maketitle

\begin{history}
\received{(Day Month Year)}
\revised{(Day Month Year)}

\accepted{(Day Month Year)}
\comby{(xxxxxxxxxx)}
\end{history}

\begin{abstract}
Quark matter is expected to exist in the interior of compact
stellar objects as neutron stars or even the more exotic strange
stars, based on the Bodmer-Witten conjecture. Bare strange quark
stars and (normal) strange quark-matter stars, those possessing a
baryon (electron-supported) crust, are hypothesized as good
candidates to
explain the properties of a set of peculiar stellar sources as the
enigmatic X-ray source RX J1856.5-3754, some pulsars as PSR
B1828-11 and PSR B1642-03, and the anomalous X-ray
pulsars and soft gamma-ray repeaters. In the
MIT bag model, quarks are treated as a degenerate Fermi gas
conf{}ined to a region of space having a vacuum energy density
$B_{bag}$ (the Bag constant). In this note, we modif{}y the  MIT
Bag Model by including the electromagnetic interaction. We also
show that this version of the MIT model implies the anisotropy of
the Bag pressure due to the presence of the magnetic f{}ield. The
equations of state of degenerate quarks gases are studied in the
presence of ultra strong magnetic f{}ields. The behavior of a
system made-up of quarks having (or not) anomalous magnetic
moment is reviewed. A structural instability is found, which is
related to the anisotropic nature of the pressures in this highly
magnetized matter. The conditions for the collapse of this system
are obtained and compared to a previous model of neutron stars
build-up on a neutron gas having anomalous magnetic moment.
\end{abstract}


\section{Introduction}    

The relation between strong magnetic f{}ields and dense matter is
a subject that have attracted so much attention recently,
especially after the observations of peculiar X-ray emission from
anomalous X-pulsars (AXPs) and low energy $\gamma$-ray radiation
from soft gamma-ray repeaters (SGRs). The central engine of these
radiations is believed to be a neutron star or 'quark star'
endowed with a magnetic f{}ield larger than $10^{13.5}$ G. The
x-ray binaries dubbed as Galactic Black Hole Candidates have also
been recently suggested as possessing a strange star as its
primary
Ref.~\refcite{lugones-horvath-02},\refcite{lugones-horvath-03},\refcite{lugones-horvath-04}.

The existence of quark stars was proposed in 1969, about f{}ive
years after than the Gell-Man prediction of quarks. Bodmer Ref.~
\refcite{Bodmer} also advanced the idea of strange quark matter in
1971. By quark matter is meant here a Fermi gas of 3A quarks which
together constitute a single color-singlet baryon with baryon
number A. It is known two types of 'quark matter': the f{}irst one
is called 'non-strange matter', which consists only of two
f{}lavor $u$ and $d$ (the 2SC model), while the second one, dubbed
'strange matter', is made-up of 3 f{}lavors $u$, $d$ and $s$.
Strange matter has the particularity that it guarantees the
equilibrium with weak interactions. In 1984 Witten Ref.~
\refcite{Witten}, and later Farhi and Jaffe Ref.~\refcite{Jaffe},
showed that for strange matter, the binding energy could be lower
than for Fe over a rather wide range of QCD parameters. Witten
Ref.~\refcite{Witten} has pointed out two possible pathways to
form strange matter: the quark hadron phase transition in the
early Universe and the conversion of neutron stars into strange
ones at extremely high densities. In this way the most stable 
compact star
might be a strange star rather than a neutron star. Neutrons and
protons are both constituted of quarks, hence as a highly dense
system of nucleons, a neutron star, especially its core, might
have been deconf{}ined to a quark (plus gluons) system under the
extraordinary conditions imposed by a a very dense nuclear matter 
and supercritical magnetic f{}ield. Because of its extremely
high density ($\geq 10^{14}$ g/cm$^3$), a neutron star might
actually be a pure 'quark star', or alternatively a neutron star
might contain a 'quark core'.

The equation of state (EOS) of both strange and non-strange
matter, and the structure and stability of stars constituted of
each of these types of matter have been studied using simplif{}ied
models
such as the MIT Bag Model Ref.~ \refcite{Jaffe} (see also Lugones
\& Horvath 2003, and the important set of actualized references
therein). The
MIT Bag Model is the most important hypothesis introduced to study
the properties of quark matter for quark stars. As such, it
considers the balance quarks as a relativistic free-particle
system conf{}ined in an impenetrable bag whose equilibrium rests
on the bag constant $B_{bag}$, i.e.,

\begin{equation}
B_{bag} = - \frac{1}{2} \frac{\partial}{\partial r}
(\bar{\psi}{\psi}) \label{bag-constant} ,
\end{equation}

\noindent where $\psi$ def{}ines the quark f{}ield. In other
words, the balance pressure on the bag surface stems from the
outward Fermi gas pressure counterbalanced by the inward (vacuum) 
bag pressure, which mimics the strong interaction that hold the
quarks together.

On the other hand, the matter in this  condition, extremely dense,
exhibits  novel properties which are worthwhile to study. Several
authors have studied equations of state and thermodynamical
properties of these systems and all have concluded that the
physical behavior change
Ref.~\refcite{Witten}-\refcite{chaichian}: Thermodynamical
properties, dynamics, etc. In particular, in the paper
Ref.~\refcite{aurora} we study the neutron gas in presence of a
strong magnetic f{}ield with the aim to describe neutron stars. We
unveiled the anisotropic behavior of the pressures perpendicular
and parallel to the magnetic f{}ield direction. This result
implies that the 'shape' of some astrophysical object endowed with
extremely high magnetic f{}ields may become prolate once quantum
effects are taken into account. Other important conclusions of
this paper suggest that we can determine the physical conditions
for which the pressure perpendicular to the (dipole) magnetic
f{}ield could vanish, and thus for the system to be unstable to
collapse pulled down by its own gravity. It is then timely to
emphasize that when these systems are described within a classical
approach the conclusions are quite opposite, in the sense that the
pressure parallel to the $B$-f{}ield turns out to be smaller than
that perpendicular to it Ref.~\refcite{khalilov}. The star then
takes an oblate shape and the collapse may take place in the
parallel direction to the f{}ield. The shape the astrophysical
object would adopt would be f{}lat  or toroidal, i.e., similar to
a disk or torus perpendicular to the $B$-f{}ield
 Ref.~\refcite{Latimer}.

Strong magnetic f{}ields are known to exist in the interior of
compact stars. Notwithstanding, only a few attempts to study the
behavior of quark matter permeated by a magnetic f{}ield have been
performed so far Ref.~\refcite{Chakrabarty}.

The scope of this paper is to show that for 'quark matter' acted
upon by a strong magnetic f{}ield the same anisotropic behavior
obtained earlier in Ref.~\refcite{chaichian}- \refcite{aurora}
holds. For the sake of simplicity, we focus here on the case of
non-strange quark matter, as a degenerate Fermi gas of quarks.
Nonetheless, our treatment and conclusions prove to be independent
of the specif{}ic model being used. As could be seen later, it is
possible (self-consistently) to use the MIT bag model in presence
of a  magnetic f{}ield and the result leads to obtain an
anisotropic Bag-pressure that depends on the direction of the
magnetic f{}ield. The equation (1) is modif{}ied because spinors
interact with the magnetic f{}ield.

To study the degenerate quark gas in presence of a magnetic
f{}ield we follow two approaches: in the f{}irst one, one
considers the quark f{}ield interacting with magnetic f{}ield via
its charge. In the second one, it is considered that quarks have
anomalous magnetic moment. At the end, we obtain little
differences between both approaches which are related to the
stability of the system. When we study a model which takes into
account the anomalous  magnetic moment, then the system becomes
more stable; in the sense that for a f{}ixed density it is
necessary to have a more stronger magnetic f{}ield in order to
bring the system into instability.

Chakrabarty et al. Ref.~\refcite{Chakrabarty} analyzed the
conditions for stability of nucleonic bags, e.g., neutrons and
protons, and obtained the instability condition for these
particles inasmuch the same way as we do here for a quark gas.
Although  we agree with their statement that nucleons immerse in a
huge magnetic f{}ield become unstable, we have some criticisms to
their approach to the issue.  The f{}irst one is related to the
use of our thermodynamical arguments for nucleons since these are
3 particle systems (it is not completely satisfactory to use
quantum statistical physics to study a few body system). The
second one is that in the extremely degenerate case  the
anisotropic behavior of pressures obtained in this paper
Eq.(5)-(6) is such that the transverse pressure goes to zero when
the magnetic f{}ield grows while the parallel pressure grows with
the f{}ield strength. Counter to this view, those authors claimed
just the opposite (such a behavior only occurs in the classical
case).

This paper is organized in the following way. In section 2 we
describe shortly the general conditions of two-f{}lavor,
non-strange quark matter of the MIT model in presence of a
magnetic f{}ield. We also rewrite the energy-momentum tensor and
state its relation to the EOS of the system. In section 3 we
study the system of a degenerate quark gas by f{}irstly including
the anomalous magnetic moment, and after by ignoring it. We also
include a comparison of our results here with the model of the
neutron gas with anomalous magnetic moment as a toy model of
neutron stars Ref.~\refcite{aurora}. Section 4 gives the
conclusions.

\section{Bag Model in presence of a Magnetic F{}ield}
To study quark matter we have to use QCD. However, for our purpose
here is enough to use the MIT-Bag Model Ref.~\refcite{Jaffe},
which in a phenomenologically way mimics the strong interaction.
We start by introducing a couple of basic parameters: $\alpha_{c}$
the coupling constant, and $B_{bag}$ as the vacuum pressure in the
MIT Bag Model.

Nevetheless, in this paper  it is enough for us to work with a
two-f{}lavor quark matter despite it is not in equilibrium with
the weak interaction. We essentially want to show that the
presence of a magnetic f{}ield in a quark matter ensemble forces
the appearance of an instability which is of similar nature to
that we have previously found for both the electron and the
neutron gas Ref.~\refcite{chaichian}-\refcite{aurora}.

The  Lagrangian density including the MIT $B_{bag}$ Model in
presence of a magnetic f{}ield should be written as

\begin{equation}
\mathcal{L}_{Bag}=[{\overline{\psi}(i\gamma_{\mu}(\partial^{\mu}-ie_qA^
{\mu})-m_q)\psi}
-1/4\mathcal{F}_{\mu\nu}\mathcal{F}^{\mu\nu}-B_{Bag}
]\theta_{\nu}(x)-\frac{1}{2}\overline{\psi}\psi\Delta_{s}\label{Lagrangeana}
\end{equation}

\noindent where  $\psi$ is the wave function of quarks $m_{q}$
represents the current quark masses and $e_q$ corresponds to the
quark charges. $B_{bag}$ is the bag constant, and the parameter 
$\theta_{\nu}$ taking the values:
$\theta_{\nu} = 1$ inside the Bag, while outside it takes on
$\theta_{\nu} = 0$. $\partial \theta_{\nu}/\partial
x^{\nu}=n_{\nu}\Delta_{s}$, with $\Delta_{s}$ being the surface
$\delta$-function and $n_{\nu}$ is a space-like unit vector
normal to the surface, and $\mathcal{F}_{\mu\nu}$ def{}ines the
electromagnetic (Maxwell) tensor.

The total energy-stress tensor has the form

\begin{equation}
T_{\mu\nu} = \frac{\partial \mathcal{L}_{Bag}}{\delta a_{i,\mu}}
a_{i,\nu} -\delta_{\mu\nu} \mathcal{L}_{Bag} ,
\end{equation}

\noindent where the index $i$ denotes the f{}ields (either
fermions or vector components).  In this case $a_{i}$ refers
to the $\psi$ and $A_{\mu}$ f{}ields Ref.~\refcite{eliza}.

In that way $T_{\mu\nu}$ has the form

\begin{equation}
T_{\mu\nu}= \left(\frac{1}{2}[i\overline{\psi}
\gamma^\mu\partial^\nu\psi - i \partial^\mu \overline{\psi}
\gamma^\nu\psi] - \frac{\partial\mathcal{L}_{Bag}}{\partial
A_{l,\mu}}A_{l}^{,\nu}-B_{bag}\right) \theta_{v} -
g^{\mu\nu}\mathcal{L}_{Bag} . \label{Tensor}
\end{equation}

Using the energy-momentum conservation written as
$T^{\beta\nu}_{,\, \, \, \, {\beta}} = 0$ we obtain

\begin{equation}
 B_{bag} \Delta_{s} n^{\nu} + \left(\frac{i}{2} \left[\overline{\psi}
\gamma^{\mu} \partial^{\nu}\psi - \partial^{\nu} 
\overline{\psi}\gamma^{\mu}\psi \right]
- \frac{\partial\mathcal{L}_{Bag}}{\partial 
A_{l,\mu}}A_{l}^{,\nu}\right) n_{\mu}
\Delta_{s}= 0 ,
\end{equation}

and

\begin{equation}
B_{bag}n^{\nu}=\frac{1}{2}\frac{\partial}{\partial
x_{\nu}}{\overline{\psi}}\psi+n_{\mu}\frac{\partial\mathcal{L}_{Bag}}
{\partial A_{l,\mu}}A_{l}^{,\nu}\label{Bag1} .
\end{equation}

This equation is the pressure balance equation. Here we take the
normalization condition $n^{\mu}n_{\mu} = -1$ and considering a
pure constant magnetic f{}ield with $A_{\mu} = \frac{B}{2}
[-x_2,x_1, 0,0]$, and the gauge derivative $\partial_{\mu} 
\rightarrow \partial_{\mu} - ie_{q}A_{\mu}$.

In Ref.~\refcite{Tesis} was proved that the last term in
Eq.(\ref{Bag1}) is proportional to $B^2$ and it is absorbed by a
renormalization process. Thus, only the f{}irst term of
(\ref{Bag1}) remains. More detailed calculations will appear in
Ref.~\refcite{Miguel}. If we take $n^\mu = (0,x_i), i = 1, 2, 3$
then $B_{bag}$ takes the form

\begin{equation}
B_{bag} = - \frac{1}{2} n_{i} \left(\frac{\partial}{\partial x_{i}}
{\overline{\psi}(x)}\psi(x)\right)
 \label{Bag}
\end{equation}

\noindent and therefore $B_{bag}$ in the presence of a magnetic
f{}ield has an anisotropic form which depends on the $B$-direction
in space, here related to the $n_i$ direction.

\subsection{Quark matter EOS in presence of magnetic f{}ield}







This section  is  devoted to obtain the equation of state of
magnetized quark matter which starting point is the statistical
average of energy-momentum tensor, Eq.(\ref{Tensor}).
To do that we use standard methods of finite temperature quantum
field theory Ref.~\refcite{Fradkin}.

The calculation of $\mathcal T_{\mu\nu}=< T_{\mu\nu}>_s$ formally
means to replace the Lagrangian by the thermodynamical potential
$\Omega$, which then is given by the expression

\begin{equation}
\Omega = - \frac{1}{\beta} \ln <e^{\int_0^\beta dx_4 \int d^3x
\mathcal{L}_{Bag}(x_4,x)}>\label{omega}
\end{equation}

\noindent where the four-components of $x_\mu$ have the form $x_\mu 
\equiv (x_4,\overrightarrow{x})\equiv(ict,\overrightarrow{x}))$, and
$\beta=\frac{1}{\kappa T}$, with  $\kappa$ the Boltzman constant and 
T is the temperature.

Let us note that in the context of the modified MIT Bag model to
describe magnetized quark matter, only
what happens inside the bag is interesting, so we will put
$\theta_\nu = 1$ in both the Lagrangian and thermodynamical 
potential.


The calculation of the thermodynamical potential is presented in
next section. Starting from it we can calculate the energy
momentum tensor. As we see below, the structure of $\mathcal
T_{\mu\nu}$ remains model independent. Similar structure was  
obtained in  Ref.~\refcite{Tesis}, \refcite{aurora},
\refcite{chaichian} where
the energy-momentum tensor was derived from QED and electroweak
theory. The thermodynamical potential contains information about
the matter and electromagnetic interactions, and the electromagnetic
tensor also appears in the structure.


Let us note that  these steps  have been followed in details in
Ref.~\refcite{chaichian}-\refcite{aurora} for degenerate gases of
electrons and neutrons. For that reason in this note we do not
devote time in deriving the energy momentum tensor, which can be
deduced doing a careful extrapolation of the previous papers in
Ref.~\refcite{chaichian}-\refcite{aurora}. The expression of
${\cal T}_{\mu \nu }$ has the form

\begin{equation}
{\cal T}_{\mu \nu }  = \left(T \frac{\partial \Omega}{\partial T} + \sum \mu_i
\frac{\partial \Omega }{\partial \mu_i}\right)
\delta_{4\mu}\delta_{4\nu} + 4 F_{\mu}^{\lambda}F_{\nu \lambda }
\frac{\partial \Omega}{\partial F^2} - \delta_{\mu \nu }\Omega,
\label{TPS}
\end{equation}

\noindent where for the sake of simplicity we assume

\begin{equation}
 \Omega = \Omega_q - B_{bag} ,
 \end{equation}


\noindent and $\Omega_q$ is the thermodynamical potential of the
magnetized degenerate quark gas, (obviously independent of the
$B_{bag}$ constant), and $q$ denotes the species of quarks.

Let us recall that the expression of ${\cal T}_{\mu \nu }$ is
general and would be used to take into account the electric and
magnetic interactions. The scope of this paper is to analyze only
magnetic effects, so in what follow we consider the magnetic field 
pointing in the direction $x_3$.

All off-diagonal components of this tensor vanish. The diagonal
components of the tensor $\mathcal{T}_{\mu\nu}$ corresponds to the
energy density and the pressures, albeit the last ones are
anisotropic due to the magnetic f{}ield. The tensor can be written
as follows

\begin{equation} {\cal T}_{\mu\nu} = \left(
\begin{array}{llll}
U & 0&0&0 \\
0& P_{\perp}&0&0\\
0&0&P_{\perp}&0\\
 0&0&0&P_{3}
\end{array}
\right) ,
\hspace{1 cm} P_{\perp}=P_{3}-{\mathcal M}B , \hspace{0.5 cm}
P_{3} = - \Omega . \label{presiones}
\end{equation}

\noindent Equations (\ref{presiones}) show that $P_{\perp} \leq
P_{3}$ if  the magnetization is a positive quantity, and thus the
behavior of the gas is paramagnetic. In the next section we
calculate explicitly the magnetization and the pressures. We can
prove that the magnetization is a positive quantity.

The condition of quantum-magnetically-induced transverse collapse:
 $P_{\perp} = 0$,
discussed in Ref.~\refcite{chaichian}-\refcite{aurora}, implies for
the magnetically modified MIT Bag model, the relation

\begin{equation}
B_{bag}^{\perp} = - \Omega_q - \mathcal{M}B.
\label{bag1}
\end{equation}

However, this condition would be enforced also for the Bag
collapse, together with

\begin{equation}
B_{bag}^{\parallel} = - \Omega_q .
\label{bag2}
\end{equation}

This would mean an anisotropic bag pressure (which is otherwise
expected from (\ref{Bag}) since the nucleon is deformed by the
magnetic f{}ield), and in place of (\ref{omega}) we would have
$\Omega \delta_{\mu\nu} \rightarrow \delta_{\mu\nu}\Omega_q -
B_{\mu\nu}^{Bag}$. Notice, nonetheless, that in the absence of the
magnetic f{}ield the pressure becomes isotropic and we recover the
condition of stability (or instability) of the Bag model given by
$B_{bag} = - \Omega_q$.

Thus, we conclude that a magnetic f{}ield brings in an instability
to the system, but it is compatible with the Bag Model if we
consider that the Bag is not isotropic and the nucleons are
deformed in a prolate-shape.

\section{Degenerate quark gas with/without anomalous magnetic moment(AMM)}

Our aim in this section is to discuss the collapse described in the 
previous section.  We derive the thermodynamical quantities of the
degenerate quark gas in presence of an ultra strong magnetic
f{}ield. In particular we calculate the thermodynamical potential
and magnetization to evaluate the anisotropic pressures Eq.
(\ref{presiones}).

In general, the quark spectrum in the external f{}ield $B$ has the
form

\begin{equation}
E_{q} = \sqrt{p_3^2 + \left( \sqrt{2e_qBn + m_{q}^2} + \eta Q_{q}B
\right)^2},
\end{equation}

\noindent where  $e_q$ is the quark's charge, $m_{q}$ represents the
quark masses, $\eta=\pm$ are the eigenvalues corresponding to two
orientations of magnetic moment (parallel and antiparallel), and
$Q_{q}$ anomalous magnetic moment of quarks. We also define $y_{q} 
= Q_{q}/m_{q}$, $b_{q} = 2e_{q}/m_{q}^2$ as
relative quantities and $x_{q}=\mu _{q}/m_{q}$, $ g_{q}(x_{q},B,n)
= \sqrt{x_{q}^2-h_{q}(B,n)^2}$ and $h_{q}(B,n) = \sqrt{b_{q}Bn +
1} + \eta y_{q}B$ as dimensionless ones.\footnote{Thermodynamical
quantities have been written here in the cgs system and $m_q$ has
energy dimension.}

For the sake of simplicity we study here nonstrange matter but the
conclusions for strange matter are esentially the same. In this case:
$Q_u = 1.82\mu_{N}$, $ Q_d = -0.9\mu_{N}$,  being $\mu_{N}$
the nuclear magneton $\mu_{N}=hc/m_{n}$.
For quarks the critical $B$-f{}ield at which the coupling energy
of its magnetic moment equals the rest-energy   is for $u$-quarks
$B^c_{u} = 6.3 \times 10^{15}$~G and for $d$ $B^c_{d} = 1.2 \times
10^{16}$~G.

Thus, for the quarks thermodynamical  potential Eq. (\ref{omega}) we
get

\begin{equation} 
\Omega = \sum \Omega_q, 
\end{equation}

with 

\begin{equation}
\Omega_{q} = - \Omega^0_{q}B \sum_n^{n_{max}}\sum_{\pm \eta}
\left[x_{q}g_{q} - h_{q}^2 \ln\frac{x_{q} + g_{q}}{h_{q}} \right],
\end{equation}

\noindent where $\Omega^0_{q} = \frac{e_q m_{q}^2 B}{4\pi ^2
(\hbar c)^2}$, and  the sum over $n$ represents the sum over
Landau levels up to $n_{max}$ given by the expression

\begin{equation}
 n_{max} = I \left(\frac{(x_{q} - \eta y_{q}B)^2 -
1}{b_{q}B} \right)\label{Landau} 
\end{equation}

\noindent where $I$ stands for the integer part function. The
magnetization is then given as

\begin{equation}
{\mathcal M} = \sum {\mathcal M}_q ,
\end{equation}

with

\begin{equation}
 {\mathcal M}_{q}  = {\mathcal M}^0_{q} \sum_n \sum_{\pm
\eta} \left\{g_{q}  x_{q} -  \left(x_{q}^2 - 2h_{q} B
\left[\frac{b_{q}n}{2\sqrt{b_{q}Bn+1}}  +  \eta y_{q}
\right]\right) \ln\frac{x_{q} + g_{q}}{h_{q}}\right\},
\label{magnetizacion}
\end{equation}

\noindent with ${\mathcal M}^0_{q} = \frac{e_q m_{q}^2}{4\pi
^2(\hbar c)^2}$ . The expression (\ref{magnetizacion}) is always 
a positive quantity because the first term $g_{q} x_{q}$ is greater
than the second one.

The density of particles has the form

\begin{equation}
N =\sum_q N_q
\end{equation}

with

\begin{equation}
N_{q} = N^0_{q} \left(\frac{B}{B^c_{q}}\right) \sum_n \sum_{\pm\eta}
g_{q}(x_{q},B,n),
\end{equation}

\noindent where $N_{q}^0 =  \frac{m_{q}^3}{(4\pi ^2(\hbar c)^3)}$,
and $B^c_{q} = \frac{m_{u,d}^2}{(e_q\hbar c)}$.

The charge neutrality in this case is given by the expression

\begin{equation} N_d = 2N_u .\label{densidad}
\end{equation}

\noindent

The presence of the magnetic field introduces restrictions to
given values of the chemical potential of the species of 
quarks. Using  Eq. (\ref{Landau}) and Eq. (\ref{densidad}) we can
compute numerically the relationship between the chemical
potentials $\mu_d$ and $\mu_u$ and their corresponding
dimensionless quantity $x_q$.



By bringing the relations for the pressures given in
(\ref{presiones}) to the expressions for the Bag pressures we
obtain for the anisotropic pressures the following expressions

\begin{equation}
B_{bag}^{\parallel}=P_{\parallel} = - \frac{e_q m_{q}^2 B}{4\pi
^2\hbar c}\sum_n\sum_{\pm \eta} \left[x_{q}g_{q} -
h_{q}^2\ln\frac{x_{q} + g_{q}}{h_{q}} \right] ,
\end{equation}

\begin{equation}
B_{bag}^{\perp}=P_{\perp} = \frac{2 e_q m_{q}^2 B^2}{\pi^2(\hbar
c)^2} \sum_n \sum_{\pm\eta} \left(2h_{q} \left[ \frac{b_{q}
n}{2\sqrt{b_{q} Bn + 1}} + \eta y_{q} \right] \ln\frac{x_{q} +
g_{q}}{h_{q}}\right) .
\end{equation}

Let us remark that we can also study a model of a degenerate quark
gas without taking into account the quark magnetic moment. This
means to make $y_{q} = 0$ in all the equations above. We recover
in that case the expressions for all the thermodynamical
quantities.

In Fig.1 is shown the phenomenon of anisotropy of the pressures.
In this case, we observe that the limiting case $P_{\perp} = 0$ is
possible to achieve for typical values of the density $N =
10^{39}$~cm$^{-3}$ and $B$-field typical of the interior of
millisecond spinning just-born neutron stars $B \sim 10^{17}$~G.

\begin{figure}[th]
 \centerline{\psfig{file=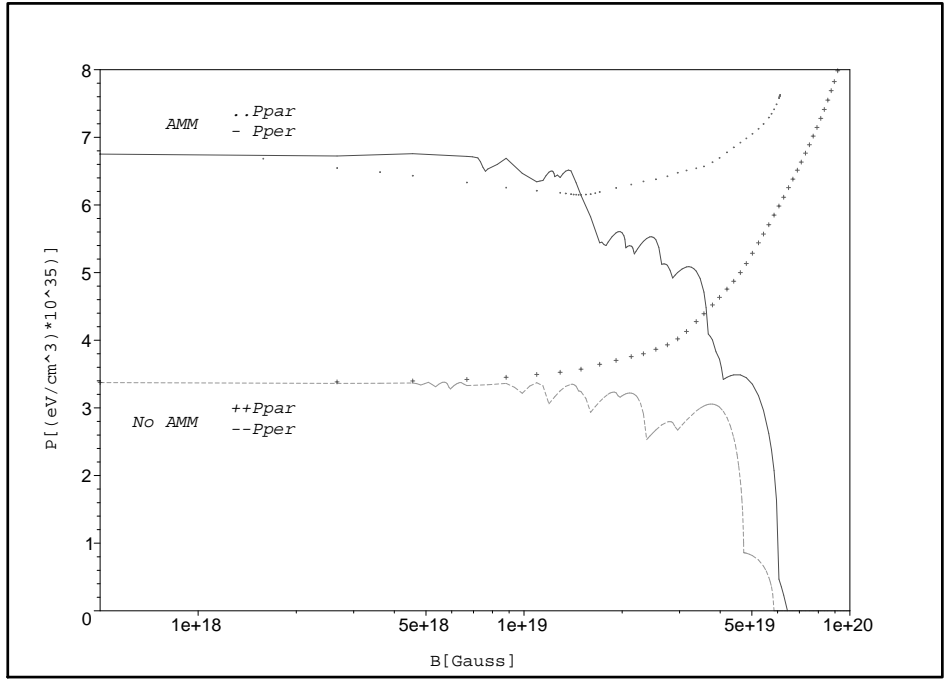,width=9cm,angle=270}}
\vspace*{8pt}
\caption{Anisotropy of pressures for the two cases studied above:
a gas of quarks $(u,d)$ having anomalous magnetic moment, and 
without it. The first being more stable as the pressure $P_{\perp}$
goes to zero for a large magnetic field.}
\label{fig1}
\end{figure}

The relation between the particle density and magnetic f{}ield
strength that fulf{}ills the condition $P_{\perp} = 0$ is the
following

\begin{equation}
N_{q}(B) = \sqrt{2} N^0_{q}y_{q} \left(\frac{B^2}{B^c_{q}}\right) .
\end{equation}


\begin{figure}[th]
\centerline{\psfig{file=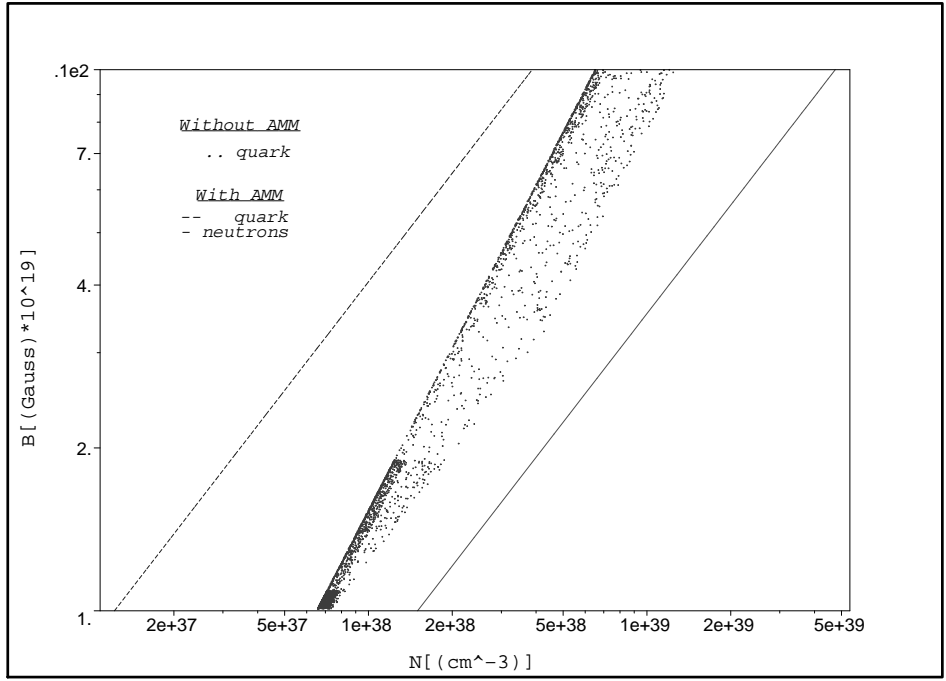,width=9cm,angle=270}}
\vspace*{8pt}
\caption{The condition $P_{\perp}= 0$ for the two quark models
with/without anomalous magnetic moment and for a model of a
neutron gas as discussed in Ref.8}
\label{fig2}
\end{figure}

In Fig.2 we present our result for the degenerate quark gas with
and without anomalous magnetic moment. We also plot the curve
obtained by using a model of a neutron gas having anomalous
magnetic moment. We conclude that the most stable conf{}iguration
can be reached for models of quark stars. In other words, the
condition of $P_{\perp} = 0$ is obtained for a lower value of the
magnetic f{}ield in the case of a neutron gas as compared to the
quark gas. In the same figure, one can also see that a model of a
degenerate quark gas with anomalous magnetic moment gives a more
wide region of stability than both the neutron gas and the quark
gas models which have no anomalous magnetic moment. In other
words, and as expected, the degenerate quark gas with anomalous
magnetic moment is the more stable conf{}iguration.

\section{Conclusions}
We have explored the behavior of a quark gas in the presence of
extremely 
large magnetic f{}ields $B\sim m_{q}^2/e_{q}$. We used a version
of the MIT Bag Model which includes the electromagnetic interaction
between quarks. From it we verify that the $B_{bag}$ can be
replaced by an anisotropic tensor. We confirm that the
instability due to the strong magnetic f{}ield discussed in
Ref.~\refcite{chaichian}-\refcite{aurora} is present also in this
case. Finally, we show that the degenerate quark gas with
anomalous magnetic moment is more stable than the quark gas
without it.

\section*{Acknowledgements}

The work of A.P.M and H.P.R have been supported by
\emph{Ministerio de Ciencia Tecnolog\'{\i}a y Medio Ambiente}
under the grant CB0407. H.J.M.C. is a fellow of the
\textit{Funda\c{c}\~{a}o de Amparo \`{a} Pesquisa do Estado do Rio
de Janeiro} (FAPERJ), Brazil. He also acknowledges the Caribbean
 Network for ICAC-ICTP for its support. A.P.M. is Associate
 Researcher at ICTP and she would like to thank Prof. J. E. Horvath
 for some useful comments about this manuscript. A.P.M thank Prof. 
A. Cabo for useful discussions.

\end{document}